# Super-Resolution Radiography by Mechanical Supersampling and Model-Based Iterative Reconstruction using High-Z Photon-Counting Detectors

Moritz Weigt[1] and Karl Hugo E. Helin[1], Nicolas Karl Fix[1], Peter Bronsert[2],[3], Frank Goldschmidtböing[4], Thomas Thuering[5], Matthias Habl[5], Spyridon Gkoumas[5], Thomas Stein[6], Paul Martin Fleïng[1], Thomas Billoud[1], Jakob Neubauer[6], Dominik von Elverfeldt[1], Martin Peter Pichotka[1]

[1] Division of Medical Physics, Department of Diagnostic and Interventional Radiology, University Medical Center Freiburg, Faculty of Medicine, University of Freiburg, Freiburg, Germany

[2] Institute of Surgical Pathology, University Medical Center Freiburg, Breisacher Straße 115A, 79106 Freiburg, Germany

[3] Core Facility Histopathology and Digital Pathology Freiburg, University Medical Center Freiburg, Freiburg, Germany

[4] Laboratory for Design of Microsystems, Department of Microsystems Engineering—IMTEK, University of Freiburg, Georges-Köhler-Allee 102, 79110 Freiburg, Germany

[5] DECTRIS AG, Switzerland

[6] Department of Diagnostic and Interventional Radiology, University Medical Center Freiburg, Faculty of Medicine, University of Freiburg, Freiburg, Germany

*Abstract*

This study presents a practical and dose-efficient strategy for resolution enhancement in planar radiography, based on mechanically supersampled acquisition with high-Z photon-counting detectors (PCDs). Unlike prior event-based or cluster methods, our approach operates in true photon-counting mode and supports clinical flux rates. Using detector trajectories spanning multiple pixels and image registration-based shift estimation, we achieve sub-pixel sampling without requiring mechanical precision, while also compensating for motion and geometric instabilities.

An iterative reconstruction framework based on Maximum Likelihood Expectation Maximization (MLEM) with a distance-driven ray model further enhances resolution and noise robustness. Long-range supersampling additionally mitigates pixel defects and spectral inhomogeneities inherent to high-Z detectors.

Phantom studies demonstrate substantial resolution improvement and image uniformity. In comparison with a clinical mammography system, the method reveals sharper detail and more homogeneous contrast at comparable or reduced dose. The resolution gain also reduces the need for geometric magnification, enabling smaller and more cost-effective PCDs.

These results establish mechanically supersampled radiography as a clinically viable approach for micron-scale imaging, with strong potential for digital mammography and other high-resolution applications and with scan times compatible with clinical workflow.

*Index Terms*— Photon-counting detectors, super-resolution imaging, high-Z PCDs, super-sampled radiography, digital mammography.

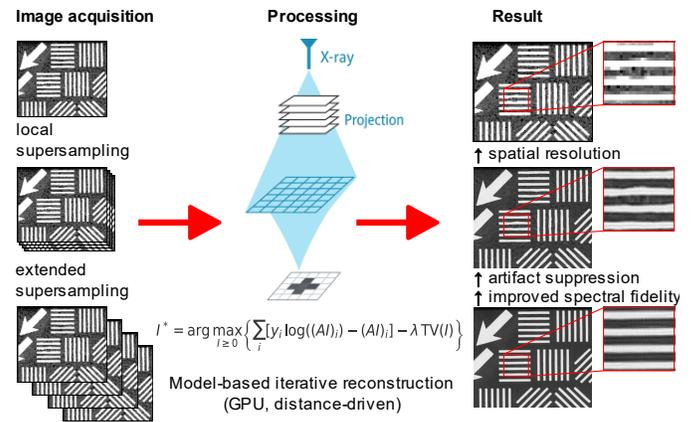

Fig. 1.  Schematic of the proposed imaging strategy. Short range supersampling and consecutive iterative reconstruction results in spatial resolution enhancement, whereas long range supersampling additionally reduces geometrical distortions and improves spectral image fidelity.

## I. INTRODUCTION

Photon-counting detectors (PCDs) offer significant benefits over energy-integrating detectors (EIDs), particularly in terms of spatial resolution, contrast, and dose-efficiency. Unlike EIDs, where light spread in scintillators introduces signal blur and necessitates design compromises such as septa or columnar structures (which reduce the fill factor), PCDs collect charge carriers directly from a high-Z semiconductor sensor. An applied electric field enables directed signal propagation, minimizing lateral dispersion and allowing for much sharper spatial localization. In addition, PCDs inherently suppress thermal and electronic noise via energy thresholding, which improves low-dose performance — a key factor in screening applications such as digital mammography.

This work was supported by the German Research Foundation (DFG, proj. no. 534069778). The authors thank DECTRIS AG of Baden, Switzerland for the loan of the PCD detectors employed in this study. The authors Thomas Thuering, Matthias Habl, Spyridon Gkoumas are employees of DECTRIS. The authors thank the Core Facility AMIRCF (DFG-RIsources N° RI_00052) for support in logistics and instrumentation.

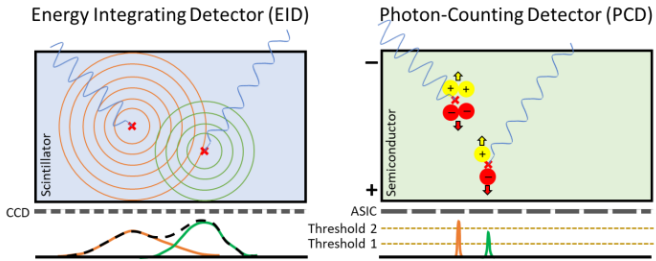

Fig. 2. Schematic of signal propagation in a scintillator-based energy-integrating detector (EID, left) and a photon-counting detector (PCD, right). Gray dashed lines indicate pixel pitch; yellow lines represent energy thresholds in the PCD. Signals exceeding these thresholds increment corresponding counters. High voltage across the PCD sensor ensures directed charge carrier propagation and therefore high contrast.

In PCD architectures like the Medipix3, [1] inter-pixel communication allows charge contributions from adjacent pixels to be assigned to the pixel with the highest signal, mitigating charge-sharing artifacts. However, spatial resolution in such detectors is still fundamentally limited by pixel pitch, and spectral performance improves with larger pixels, creating a design trade-off. Prior approaches to super-resolution using cluster-based analysis (e.g., Timepix-based centroiding [2],[3]) are limited to low-flux, sparse-event conditions, and become impractical at clinical photon flux due to dead time, communication overhead, and motion sensitivity.

To address these limitations, we propose a mechanical supersampling approach (PC-SSI) combined with iterative reconstruction. The method is compatible with clinical photon flux, operates in frame-based mode, and preserves the benefits of PCDs while achieving sub-pixel resolution. We introduce a supersampled acquisition strategy using detector motion, which avoids parallax artifacts associated with source steering. By designing sampling trajectories that span multiple pixels and avoid known sensor defects, the method also mitigates the influence of gaps between readout chips (the technical term is Application-Specific Integrated Circuit, or ASIC) and material-related blind spots, which are common to dose-efficient high-Z PCD sensor materials.[4]–[6] Long-range supersampling further averages out spectral and geometric inhomogeneities while improving overall image homogeneity.

Our reconstruction pipeline employs motion-corrected supersampling, a distance-driven raytracer tailored to PCD response, and a regularized maximum likelihood expectation maximization (MLEM) algorithm to recover high-resolution images with strong noise resilience. The clinical potential of this approach is demonstrated in phantom studies using doses consistent with mammographic screening protocols. The results suggest that PC-SSI can enhance resolution without geometric magnification, reduce reliance on large-area detectors, and support the detection of small microcalcifications and subtle lesions.

## II. Materials and Methods

### A. Imaging Setup

Experiments were performed using a high-Z photon-counting GaAs detector prototype (SANTIS HR 0804 prototype, DECTRIS AG) with a 500 µm sensor and 75 µm pixel pitch. The detector was integrated into a custom-built µCT imaging platform previously validated for high-resolution tomography. [7] X-ray imaging was performed using a micro-focus tube (MXR Microbox 100, 100 kV, 150 µA, Micro X-ray Inc.), equipped with a motorized beam-hardening calibration unit featuring interchangeable filter discs (e.g. Al, Polymethylmethacrylate (PMMA), stainless steel, Cu).

### B. Reference Samples

Four phantoms were used to validate the approach:

1. QRM Micro-CT Bar Pattern Phantom with etched patterns (150 µm–5 µm).
2. A tungsten steel knife edge to measure the edge spread function.
3. A high-contrast stainless steel grid featuring geometric calibration structures for systematic parameter assessment.
4. Gammex 156 Mammographic Accreditation Phantom, scanned at variable doses, and compared to a clinical Mammomat Inspiration system.

### C. Data Processing and Calibration

Projection data were calibrated using signal-to-equivalent-thickness (STC) normalization,[8], [9] which compensates for beam hardening and pixel-to-pixel response variations. This calibration improves shift estimation accuracy by enhancing image uniformity. STC calibration is particularly useful in computed tomography, allowing efficient suppression of beam-hardening artifacts, given that the attenuation behavior of the calibration material is sufficiently close to that of the imaged object. However, STC also compensates for inter-pixel variation in the spectral domain, which also aids planar radiography.

Sub-pixel displacements between acquisitions were determined using enhanced correlation coefficient (ECC) registration with sub-pixel precision.[10] High-contrast markers were used to support robust alignment under low-dose conditions.

The reconstruction was formulated as a pseudo-laminographic inverse problem, with negligible parallax. An MLEM algorithm was used, incorporating a distance-driven ray-tracer that models PCD pixel response.[11] Additionally, the reconstruction framework offers a total variation (TV) minimization step that can be applied to suppress noise and stabilize the reconstruction.[12]–[14]

The full workflow includes:

1. Preprocessing: STC-based correction of projection data.
2. Displacement Estimation: ECC-based sub-pixel registration.
3. Ray Tracing: Distance-driven, GPU-accelerated forward model [23,28].
4. Iterative Reconstruction: MLEM with TV regularization.
5. Evaluation: Assessment of resolution, noise, and alignment quality.

### D. Dose Measurement

Dose estimates were obtained using a JT-RAD01 digital dosimeter (SIMAC Electronics). Values were calculated from exposure time alone, neglecting readout and motion delays, to provide a conservative baseline for clinical translation.

## III. RESULTS

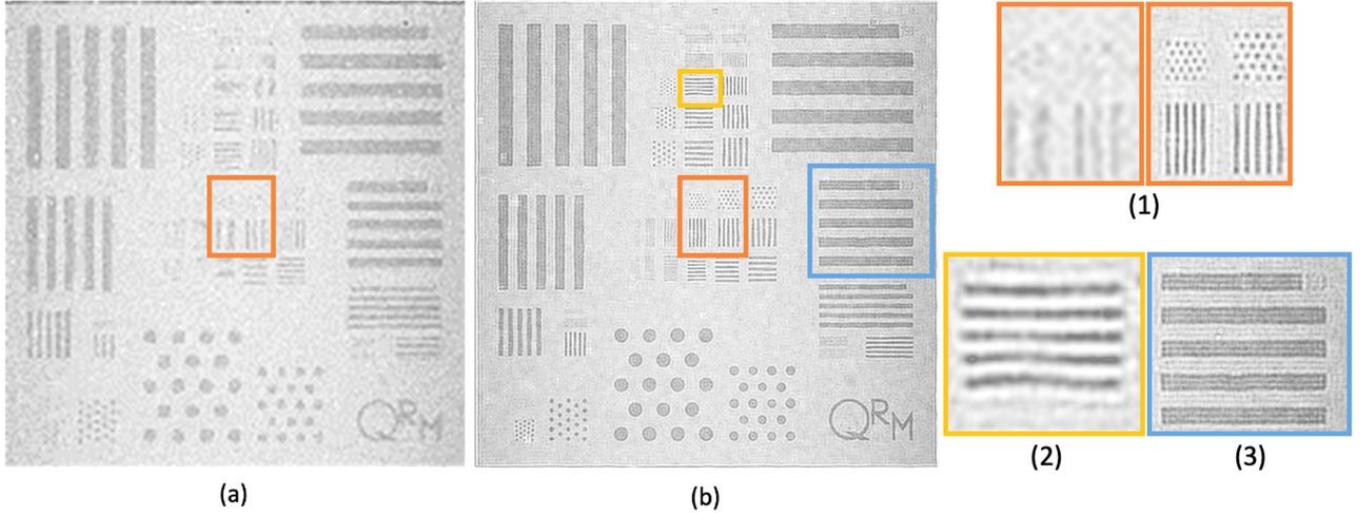

Fig. 3. Scan of micro-CT resolution phantom; prototype DECTRIS SANTIS GaAs (1280x512 px), TH40kV; Tube MXR-Microbox, 100kVp, 100muA, magnification ~2x. Comparison of phantom scans without (a) and with super-sampling (b) acquired with identical photon statistics. In both cases 21x21 frames were acquired, with and without super-sampling motion (total displacement of ~1 pixel), respectively. The effect of the spatial resolution enhancement is visible in (1), geometric distortions emerging at this increased resolution are visible in (2), whereas simultaneously emerging ringing artifacts are visible in (3).

Figure 3 illustrates the basic benefit of using supersampled PCD image acquisition in combination with the proposed iterative image reconstruction approach. The orange box (1) shows a comparison of sub-structures (20-25μm), illustrating the substantial resolution improvement by PC-SSI. The projection data were STC calibrated using PMMA filters.

Simultaneously, novel image artifacts emerge. Geometrical distortions, likely caused by sensor imhomongeneities, can be observed in the yellow box (2). The blue box (3) reveals ringing artifacts in vicinity of major structures, which we attribute to the charge sharing behavior of the detector.

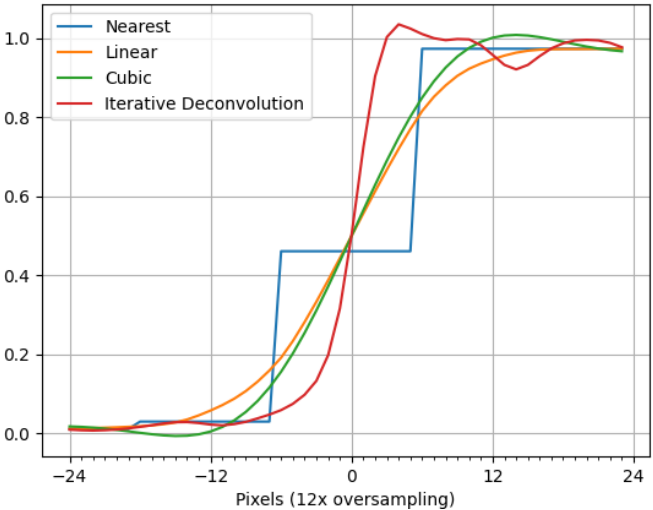

Fig. 4. Comparison of the resulting edge response function of a slanted tungsten steel edge for different image recovery approaches. Here the "nearest" response is the average per-pixel signal along the edge, "linear" and "cubic" denote an upscaled and superimposed and averaged image, using first and third order interpolation for scale and shift operations respectively. "Iterative" denotes the iterative reconstruction approach described above, assuming a pseudo-laminographic setting.

Following the image recovery approach proposed by Dreier et al. [15] we performed a comparative analysis of the edge spread function using superposition of supersampled frames by linear and cubic interpolation, and compared this to the iterative approach proposed in this paper. The "nearest" in this context represents the averaged response across the true pixel pitch of the detector. A tungsten steel knife edge was used in this measurement (see fig. 4).

The determined edge spread function (ESF) rises from 10% to 90% within 15 subpixels, or 93.75μm, in case of linear interpolation and within 5 subpixels, or 31.25μm, in case of iterative deconvolution, marking a threefold resolution improvement.

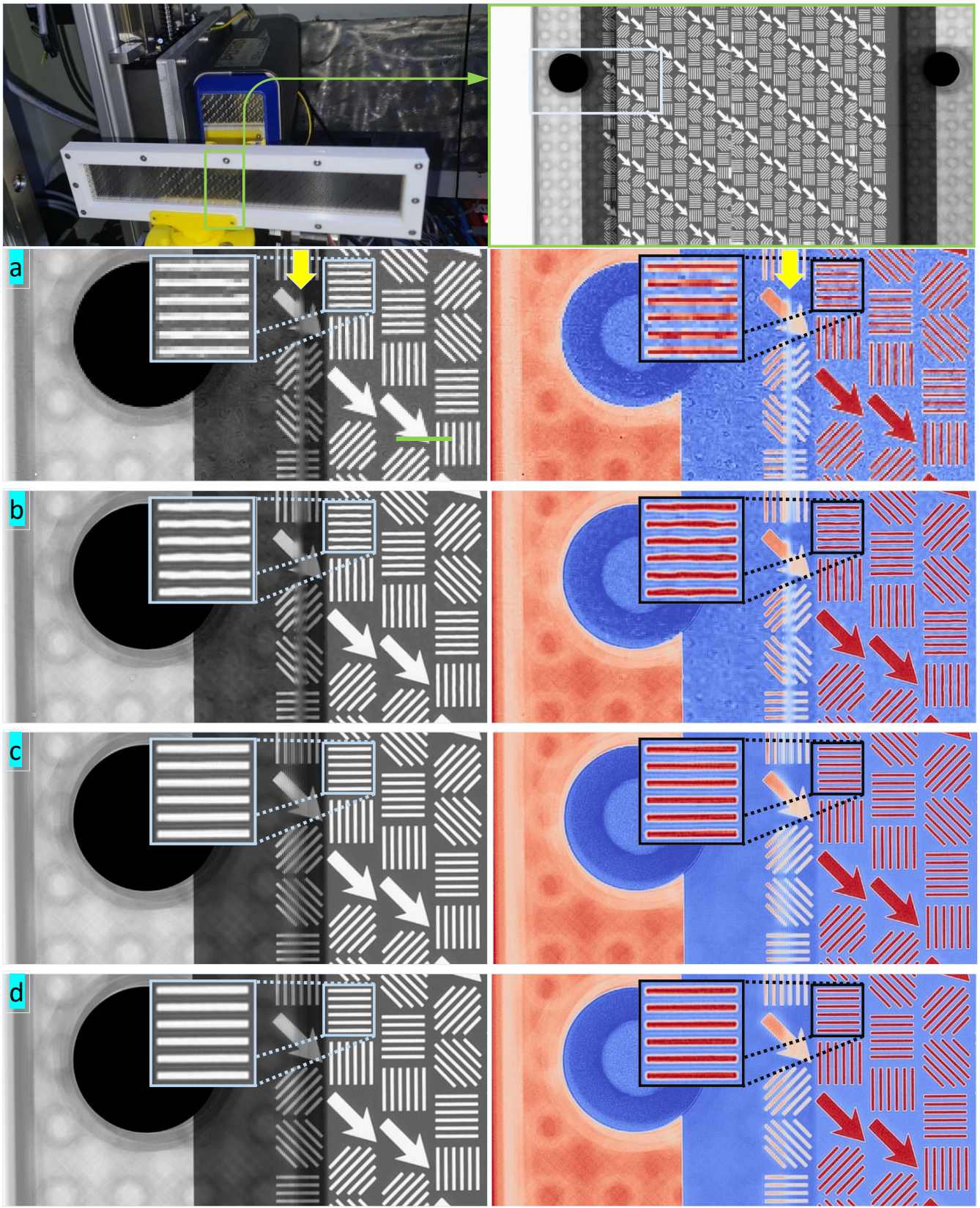

Fig. 5. Top left: Laser-cut stainless steel sheet phantom (150μm) in front of the Detector. Top right: Transmission Image (turned by 90 degrees; red box: cutout shown below). A) Native detector resolution, left: Transmission (threshold 15 keV), right: Transmission at 15 keV threshold divided by transmission at 30 keV. B) 14x14 supersampling with a stride of ~0.15 pixels. C) 11x11 supersampling with a stride of 1.1 pixels, D) 22x22 supersampling with a stride of 1.1 pixels. Cf fig. 6. The yellow arrows in subfigure *a* indicate the location of an ASIC border. Due to inconsistent response of the border pixels, spanning 2 pixels on each side of the gap, as well as of the adjacent pixel, we linearly interpolate the image over a 3-pixel region on each side of the gap.

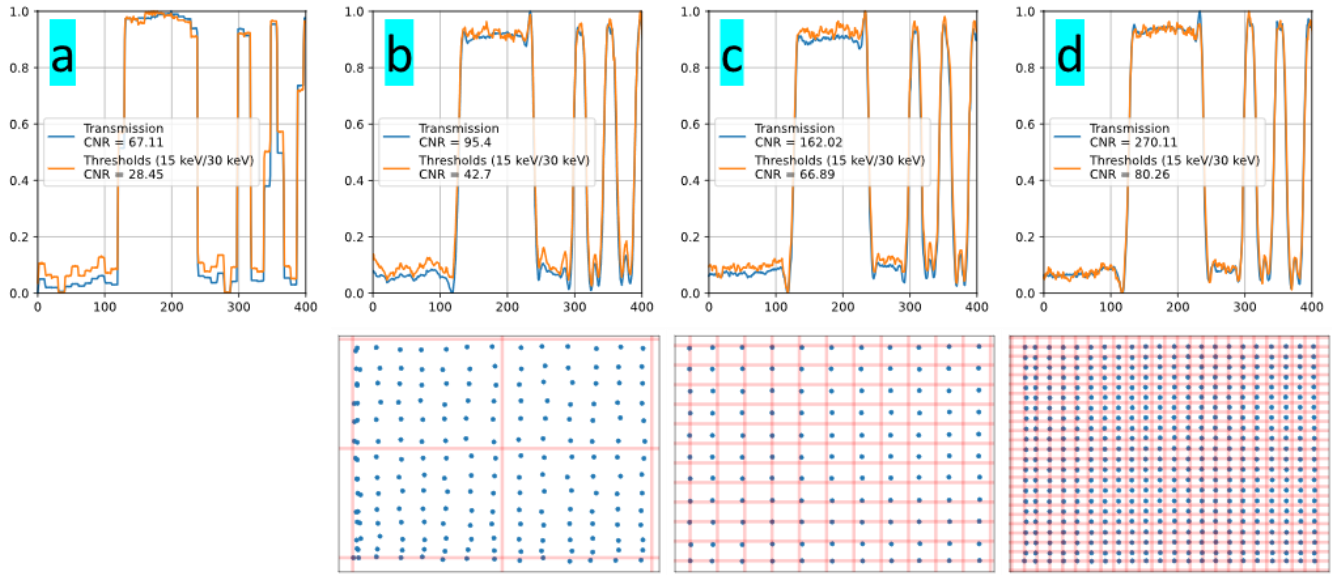

Fig. 6. Top row: Line profiles taken in the location indicated by the green line in fig. 5a and subsequent subfigures. Contrast to Noise Ratios (CNR) calculated using pixels 0:50 and 150:200. A line profile for the absorption and the spectral response are shown. Bottom row: Detector scanning trajectories corresponding to the measurements shown in figs. 5 b-d. The red boxes in the plot indicate the respective pixel boundaries.

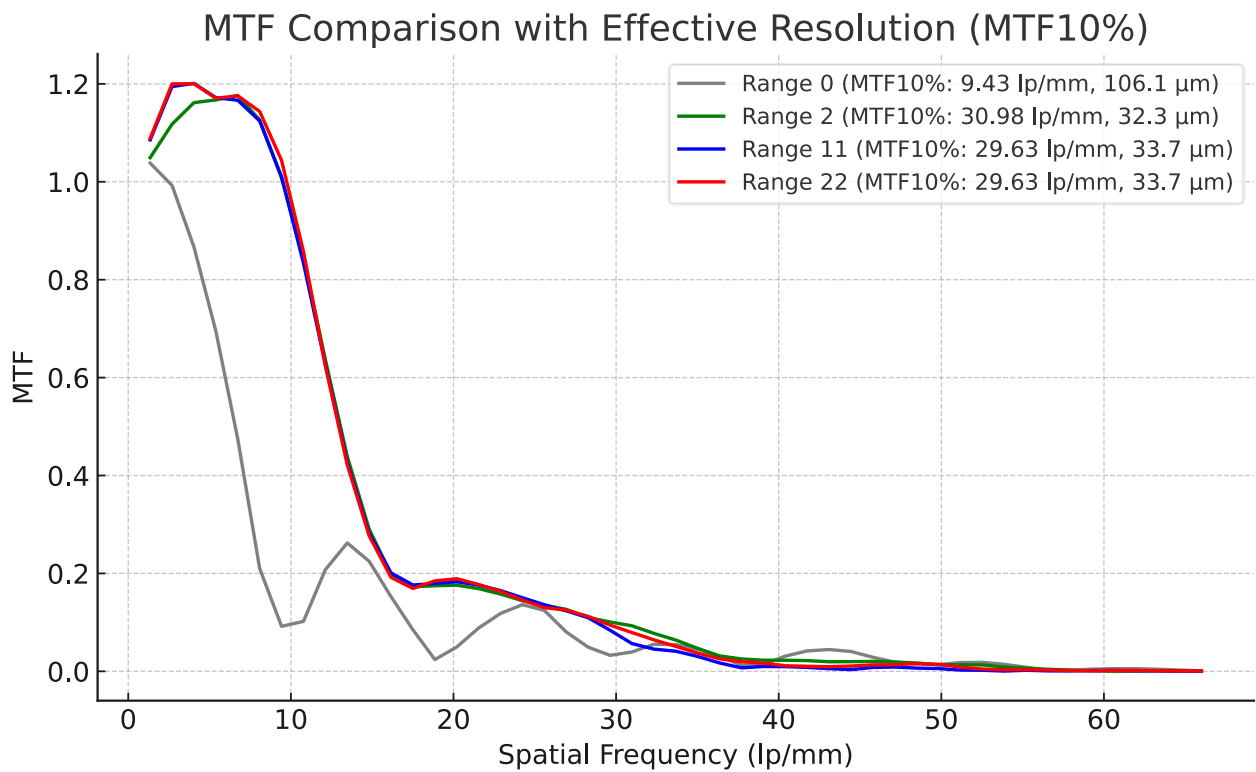

Fig. 7. Module Transfer Function (MTF) for the above approaches (calculated in the region of the arrow head indicated by the green line in figure 5a), and corresponding effective resolution. The effective resolution for each sampling trajectory in figure 5 (grey 0 / native, green 2x2, blue 11x11, red 22x22) is calculated using the MTF 10% threshold.

This work was supported by the German Research Foundation (DFG, proj. no. 534069778). The authors thank DECTRIS AG of Baden, Switzerland for the loan of the PCD detectors employed in this study. The authors Thomas Thuering, Matthias Habl, Spyridon Gkoumas are employees of DECTRIS. The authors thank the Core Facility AMIRCF (DFG-RIsources N° RI_00052) for support in logistics and instrumentation.

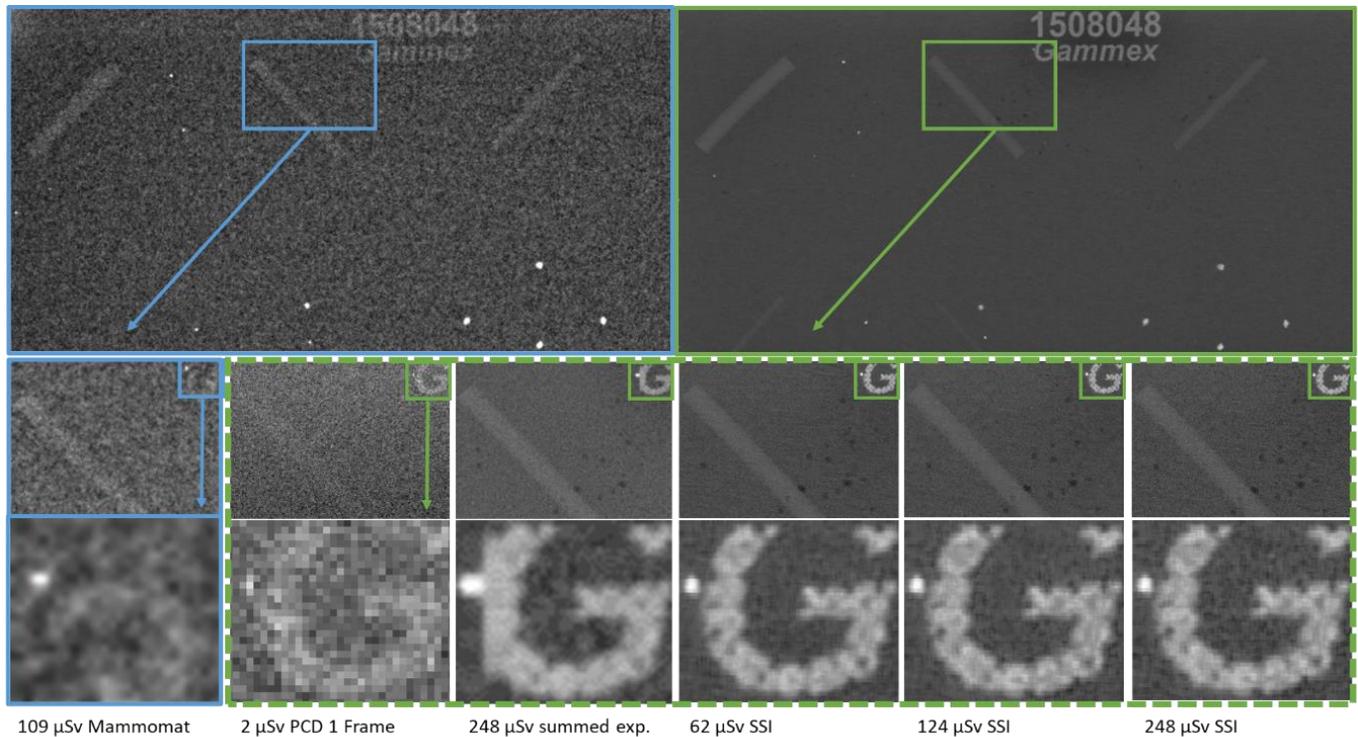

Fig. 8. Gammex ACR MAP phantom measured by 1: Mammomat Inspiration, top left (blue boxes, parameters and settings analogous to our preceding study on intraoperative imaging of breast resectates[16]) and 2: PC-SSI (green boxes) using a prototype DECTRIS SANTIS GaAs (750μm) 0804 HR prototype detector, top right. The following rows show individual features of the phantom with increasing magnification, i.e. in the second a magnification of one nylon fiber of the phantom is shown, in the third a magnification of the first letter "G" in the "Gammex" logo. From left to right the magnifications display 1: clinical mammography at 109μSv, 2: a single PCD exposure at 2μSv, 3: a summed PCD exposure at 248μSv, as well as PC-SSI images at 4-6: 62 μSv, 124 μSv and 248μSv, respectively. The supersampling matrix spanned 1 pixel (vertical) x 14 pixels (horizontal) with a 0.125 pixel step. For PC-SSI below 248μSv an interleaving pattern of sub-acquisitions from adjacent pixels was used for image reconstruction.

The sample shown in Fig. 5 was selected to evaluate resolution and spectral fidelity under realistic conditions. Its structure includes high-contrast features oriented in multiple directions, as well as regions with strong attenuation gradients (metal screw embedded in plastic). The image region selected for Figs. 5a–d includes both fiber-like elements from the 3D-printed mount and metallic components, enabling visual assessment of reconstruction quality across a range of spatial frequencies and attenuation levels.

A strong resolution enhancement is observed for all supersampled measurements. However, for short acquisition trajectories, the effect of intra pixel variations is still clearly visible, leading to grainy appearance in the absorption image and pronounced local variations in the spectral domain (see fig 5 on the right, subfigures a,b vs. c,d). Further it should be noted that with increasing range of the acquisition trajectories also image artifacts caused by sensor defects, as well as ASIC borders (the location of such a border is indicated in fig. 5 by the yellow arrows), are increasingly suppressed. The geometrical deformation artifacts of the phantom structures (likely caused by field inhomogeneities within the sensor), which only become visible due to resolution enhancement by short range PC-SSI (fig. 5b), are effectively averaged out for long range PC-SSI trajectories. The spectral line profiles (Fig. 6, top row) highlight how long-range supersampling reduces pixel-to-pixel spectral variability, effectively smoothing out detector inhomogeneities. This acts as a virtual increase in fill factor, particularly benefiting contrast agent discrimination and quantitative spectral analysis.

All measurements shown in Figs. 5–6 were acquired and processed analogously, except for the supersampling protocol. The tube was operated at 100 kV and 5 W, with a source-to-detector distance of 650 mm and a sample-to-detector distance of 75 mm. No additional filtration was applied. At a tube power of 5 W, the integrated exposure time per dataset was 4 minutes.

As noted in the caption of Fig. 5, a 3-pixel-wide region on each side of the ASIC borders was linearly interpolated during frame preprocessing. In the detector architecture used here, the outermost pixel rows and columns of each ASIC are double-width (spanning two pixel pitches). While this design allows for seamless radiographic tiling, these boundary pixels often exhibit spectral responses that deviate from bulk behavior—sometimes affecting adjacent rows or columns. To preserve the fidelity of the PC-SSI method, which relies on consistent spectral averaging, these regions were interpolated in the native data. This should not be viewed as a limitation of the detector itself, but rather as an architectural requirement for achieving gapless coverage. Importantly, the geometric alignment of submodules appears better than the pixel pitch, as evidenced by the lack of distortion at ASIC interfaces even in long-range PC-SSI reconstructions (Fig. 5d, yellow arrow).

These effects are quantitatively confirmed in Fig. 6, which shows line profiles of both the absorption signal and the spectral response across different supersampling schemes. As the sampling trajectory increases in range (bottom row), both signal homogeneity and contrast-to-noise ratio improve. The absorption profile becomes smoother and the spectral response flattens, reflecting improved averaging of inter-pixel spectral variations. Notably, the most significant Contrast to Noise Ratio (CNR) improvement is observed between 2×2 and 11×11 sampling, with diminishing returns beyond that.

The resolution enhancement observed in Fig. 7 is consistent with a near threefold improvement in effective spatial resolution, based on the Module Transfer Function (MTF) 10% threshold. Notably, the range 2 MTF exceeds those of longer trajectories at higher frequencies, reflecting a trade-off between preserved edge sharpness and artifact suppression. While short-range supersampling maintains more native sharpness, longer trajectories (11 and 22) more effectively average out geometric distortions and pixel-level inhomogeneities, resulting in improved spatial fidelity and smoother spectral response (cf. Fig. 6). To ensure consistent edge alignment across all MTF evaluations, the inclination angle (74°) was determined from the range 22 dataset, which provides the most stable and artifact-free geometric reference. Importantly, for range 0 and range 2, the MTF may be underestimated due to residual geometric distortions that displace the edge profile across lines and broaden the LSF. Unlike the MTF curves in Fig. 7, which are extracted from spatially fixed edge segments and thus susceptible to geometric distortion, the ESF in Fig. 4 was computed using line-wise alignment of edge positions, minimizing the influence of local warping and providing a cleaner estimate of intrinsic sharpness

To assess the efficacy of our approach in comparison to current clinical mammography, we performed comparative measurements of a Gammex ACR MAP using a Mammomat Inspiration system. For the PCD SSI imaging we performed STC calibration using PMMA filters.

The results shown in fig. 8 clearly indicate that PC-SSI can produce superior results compared to current clinical mammography at equivalent or better dose performance. Note that the image contrast differs to clinical mammography due to STC calibration in the PC-SSI imaging (here using PMMA) which maps count values of the detector to corresponding absorber thickness of the calibration material.

## IV. Discussion

The super-resolution strategy presented here offers a clinically viable combination of resolution enhancement, robustness, and compatibility with photon-counting mode at clinical flux levels. Unlike cluster-based approaches, our method requires no event-based processing and avoids associated flux limitations. Sub-pixel shifts are estimated through robust image registration, enabling tolerance to mechanical imprecision and patient motion.

The key innovation lies in combining mechanical supersampling with iterative reconstruction using a distance-driven projector. This enhances resolution while suppressing noise, as demonstrated by the ESF measurements in Fig. 4 and the MTF analysis in Fig. 7, where a resolution improvement by a factor of ~3 is observed over conventional interpolation. The scan of the QRM bar phantom and stainless-steel calibration grid (Figs. 3 and 5) further illustrate the visibility of fine structures down to ~20 µm. While our MLEM reconstruction framework supports regularization (e.g., total variation minimization), and preliminary experiments suggested potential benefits in noise suppression and artifact reduction, we disabled regularization for the results presented here in order to isolate and clearly demonstrate the effects of mechanical supersampling. Future work will evaluate the role of regularization strategies in further optimizing image quality under varying noise and sampling conditions.

Importantly, the results reveal a trade-off between sampling range and image artifacts. Short-range supersampling (e.g., ~1 pixel total displacement) offers the strongest resolution enhancement but leads to increased visibility of pixel-level inhomogeneities, such as:

Field distortions (visible as geometric warping),

Charge sharing artifacts (ringing),

ASIC border effects and sensor defects.

By contrast, long-range supersampling (e.g., >1 pixel stride) effectively averages out these effects across many pixels, leading to improved image homogeneity and spectral fidelity (Fig. 6, right column). This is particularly beneficial for quantitative spectral imaging and contrast agent mapping. However, the trade-off is a moderate increase in spatial blur due to averaging of shifted structures.

To overcome this, future implementations may use explicit remapping of pixel-level deformations, based on prior spectral field mapping. Because the electric field experienced by charge carriers depends on photon energy (due to variable penetration depth), correction must be energy-specific. Such spectral remapping would ideally suppress both high-frequency deformations and spectral artifacts without resorting to excessive averaging. Nonetheless, long-range sampling remains essential to suppress ASIC border artifacts and compensate for larger sensor defects.

As mentioned above, we attribute the ringing artifacts emerging at supersampled resolution mainly to charge sharing. The "Edge pixel response studies of edgeless silicon sensor technology for pixellated imaging detectors" [17] publication by Maneuski et al., where an X-ray microbeam is used to map PCD pixel response, nicely illustrates on the one hand the super-resolution potential of PCDs, with the edge response being much narrower than the pixel pitch, and the charge sharing behavior of PCDs on the other, with multiple counts attributed to a single photon in the edge region of the pixel. This behavior clearly depends on the detectors' threshold setting vs. the spectral beam composition. Indeed, in our data we observe a dependence of the ringing artifacts on the detector threshold, with the remaining parameters fixed. However, this problem exceeds the scope of this paper, and will be addressed in future work.

The proposed method supports both raster-based supersampling and "ballistic" imaging strategies. In ballistic acquisition, the detector moves linearly across the field of view with a tilted trajectory, acquiring supersampled frames in a single, continuous pass. This minimizes mechanical acceleration and acquisition time. As demonstrated in Fig. 5, a full ballistic scan with a 5 kW X-ray source could be completed in ~240 ms with detectors supporting burst readout at 2 kHz (e.g., DECTRIS EIGER2X). This duration is comparable to exposure times in current clinical mammography systems, which typically range between 100–400 ms per view [18], [19]. However, the target resolution in our method is significantly higher than that of clinical systems, and would naively require even shorter acquisition times to mitigate motion artifacts. Here, the use of image correlation not only allows accurate estimation of detector shifts, but also enables partial compensation for patient motion, further enhancing robustness. Additionally, ballistic supersampling in reverse geometry decouples spatial resolution from focal spot size, allowing the use of higher-powered X-ray sources to shorten exposure time without degrading resolution.

For practical implementation of ballistic imaging, golden-angle sampling or related non-redundant angular schemes should be considered. Due to constraints on displacement speed and exposure time, approximations to ideal radial sampling patterns may be needed to balance angular diversity and dose efficiency. This is an area for future optimization.

Among clinical applications, digital mammography (DM) stands out as particularly promising. The detection of small calcifications and subtle spiculations requires high resolution, typically below 100 µm. Recent work demonstrates that CNN-based lesion detection benefits significantly from higher spatial resolution, directly improving diagnostic performance [20]. Our approach enables this without requiring geometric magnification, thereby reducing detector area and cost. Supersampled acquisition further enables tiled scanning and region-of-interest (ROI) imaging, making large-field coverage with small detectors feasible.

Spectral fidelity also benefits from the method. Long-range supersampling combined with STC calibration and narrow energy thresholding enables inter-pixel response averaging and may allow quantitative contrast agent detection from a single projection. This is particularly relevant for contrast-enhanced mammography and other dual-energy applications.

In summary, our method provides a fast, practical, and dose-efficient solution for achieving micron-scale radiographic resolution with high-Z PCDs. It supports both raster and ballistic sampling, mitigates detector inhomogeneities, and enables new spectral imaging strategies. Ongoing work will focus on spectral field mapping, real-time motion compensation, and integration into tomographic and clinical workflows.

Beyond X-ray radiography, the PC-SSI method may also benefit laboratory-based propagation-based phase-contrast (PbPC) imaging. In such setups, refractive index variations cause slight angular deviations of X-rays, which can be visualized as interference fringes after free-space propagation. Because these refraction angles are extremely small, conventional PbPC requires either very high detector resolution or long propagation distances to convert phase shifts into measurable intensity variations.

The enhanced resolution offered by PC-SSI enables shorter propagation distances for comparable phase sensitivity, reducing the geometric footprint of PbPC systems and supporting the use of smaller detectors.

In addition, many PbPC methods exploit changes in propagation distance to discriminate between absorption and refraction signals. Alternatively, spectral imaging offers a complementary strategy: since the refractive index decrement $\delta$ is energy-dependent, the refraction angle also varies with photon energy. In principle, this enables separation of absorption and refraction components based on spectral dispersion, provided sufficient energy resolution and spectral homogeneity.

However, practical spectral PbPC applications using photon-counting detectors have been limited by inter-pixel variations in spectral response. The PC-SSI strategy addresses this by averaging over spatially shifted acquisitions, thereby reducing pixel-level spectral deviations and enhancing spectral fidelity.

These improvements may, in the long term, support clinically viable PbPC implementations in compact settings — with e.g. dental crack detection emerging as a particularly illustrative application due to its demand for high spatial resolution and sensitivity to microstructural discontinuities.

## REFERENCES

[1] R. Ballabriga, J. Alozy, G. Blaj, M. Campbell, M. Fiederle, E. Frojdh, E. H. M. Heijne, X. Llopart, M. Pichotka, S. Procz, L. Tlustos, and W. Wong, "The medipix3RX: A high resolution, zero dead-time pixel detector readout chip allowing spectroscopic imaging," *Journal of Instrumentation*, vol. 8, no. 2, Feb. 2013.

[2] A. S. Tremsin, J. V. Vallerga, O. H. W. Siegmund, J. Woods, L. E. De Long, J. T. Hastings, R. J. Koch, S. A. Morley, Y. De Chuang, and S. Roy, "Photon-counting MCP/Timepix detectors for soft X-ray imaging and spectroscopic applications," *J Synchrotron Radiat*, vol. 28, no. Pt 4, pp. 1069–1080, Jul. 2021.

[3] M. Khalil, E. S. Dreier, J. Kehres, J. Jakubek, and U. L. Olsen, "Subpixel resolution in CdTe Timepix3 pixel detectors," *J Synchrotron Radiat*, vol. 25, no. 6, pp. 1650–1657, Nov. 2018.

[4] D. Pennicard, B. Pirard, O. Tolbanov, and K. Iniewski, "Semiconductor materials for X-ray detectors," *MRS Bull*, vol. 42, no. 6, pp. 445–450, Jun. 2017.

[5] M. Fiederle, S. Procz, E. Hamann, A. Fauler, and C. Fröjdh, "Overview of GaAs und CdTe Pixel Detectors Using Medipix Electronics," *Crystal Research and Technology*, vol. 55, no. 9, Sep. 2020.

[6] P. M. Shikhaliev, S. G. Fritz, and J. W. Chapman, "Photon counting multienergy x-ray imaging: Effect of the characteristic x rays on detector performance," *Med Phys*, vol. 36, no. 11, pp. 5107–5119, 2009.

[7] M. P. Pichotka, M. Weigt, M. J. Shah, M. F. Russe, T. Stein, T. Billoud, J. Beck, J. Straehle, C. L. Schlett, D. v. Elverfeldt, and M. Reisert, "Pilot study on high-resolution radiological methods for the analysis of cerebrospinal fluid (CSF) shunt valves," *Z Med Phys*, 2023.

[8] J. Jakubek, "Data processing and image reconstruction methods for pixel detectors," *Nucl Instrum Methods Phys Res A*, vol. 576, no. 1, pp. 223–234, 2007.

[9] D. Vavrik and J. Jakubek, "Radiogram enhancement and linearization using the beam hardening correction method," *Nucl Instrum Methods Phys Res A*, vol. 607, no. 1, pp. 212–214, 2009.

[10] G. D. Evangelidis and E. Z. Psarakis, "Parametric image alignment using enhanced correlation coefficient maximization," *IEEE Trans Pattern Anal Mach Intell*, vol. 30, no. 10, pp. 1858–1865, 2008.

[11] B. De Man and S. Basu, "Distance-driven projection and backprojection in three dimensions," *Phys Med Biol*, vol. 49, no. 11, pp. 2463–2475, Jun. 2004.

[12] M. P. Pichotka, "Iterative CBCT reconstruction-algorithms for a spectroscopic Medipix-Micro-CT, https://freidok.uni-freiburg.de/data/9562," 2014.

[13] M. Pichotka, J. Jakubek, and D. Vavrik, "Spectroscopic micro-tomography of metallic-organic composites by means of photon-counting detectors," *Journal of Instrumentation*, vol. 10, no. 12, Dec. 2015.

[14] D. Vavrik, J. Jakubek, I. Kumpova, and M. Pichotka, "Dual energy CT inspection of a carbon fibre reinforced plastic composite combined with metal components," *Case Studies in Nondestructive Testing and Evaluation*, vol. 6, pp. 47–55, Nov. 2016.

[15] T. Dreier, U. Lundström, and M. Bech, "Super-resolution X-ray imaging with hybrid pixel detectors using electromagnetic source stepping," *Journal of Instrumentation*, vol. 15, no. 3, p. C03002, Mar. 2020.

[16] C. Neubauer, J. S. Yilmaz, P. Bronsert, M. Pichotka, F. Bamberg, M. Windfuhr-Blum, T. Erbes, and J. Neubauer, "Accuracy of cone-beam computed tomography, digital mammography and digital breast tomosynthesis for microcalcifications and margins to microcalcifications in breast specimens," *Sci Rep*, vol. 12, no. 1, p. 17639, Oct. 2022.

[17] D. Maneuski, R. Bates, A. Blue, C. Buttar, K. Doonan, L. Eklund, E. N. Gimenez, D. Hynds, S. Kachkanov, J. Kalliopuska, T. McMullen, V. O'Shea, N. Tartoni, R. Plackett, S. Vahanen, and K. Wraight, "Edge pixel response studies of edgeless silicon sensor technology for pixellated imaging detectors," *Journal of Instrumentation*, 2015.

[18] S. Vedantham, A. Karellas, G. R. Vijayaraghavan, and D. B. Kopans, "Digital breast tomosynthesis: State of the art1," *Radiology*, vol. 277, no. 3, pp. 663–684, Dec. 2015.

[19] I. Sechopoulos, "A review of breast tomosynthesis. Part I. The image acquisition process," *Med Phys*, vol. 40, no. 1, 2013.

[20] K. Rangarajan, A. Gupta, S. Dasgupta, U. Marri, A. K. Gupta, S. Hari, S. Banerjee, and C. Arora, "Ultra-high resolution, multi-scale, context-aware approach for detection of small cancers on mammography," *Scientific Reports 2022 12:1*, vol. 12, no. 1, pp. 1–8, Jul. 2022.

This work was supported by the German Research Foundation (DFG, proj. no. 534069778). The authors thank DECTRIS AG of Baden, Switzerland for the loan of the PCD detectors employed in this study. The authors Thomas Thuering, Matthias Habl, Spyridon Gkoumas are employees of DECTRIS. The authors thank the Core Facility AMIRCF (DFG-RIsources N° RI_00052) for support in logistics and instrumentation.